\documentclass[
twocolumn, superscriptaddress, amsmath, amssymb, aps, prx, floatfix]
{revtex4-2}

\usepackage[english]{babel}
\usepackage{graphicx}
\usepackage{hyperref}

\begin{document}

\title{4D-QENS Analysis of Correlated Ionic Conduction in SrCl\texorpdfstring{\textsubscript{2}}{2}}
\thanks{The U.S. Government retains for itself, and others acting on its behalf, a paid-up nonexclusive, irrevocable worldwide license in said article to reproduce, prepare derivative works, distribute copies to the public, and perform publicly and display publicly, by or on behalf of the Government.}

\author{Jared Coles}
 \affiliation{Materials Science Division, Argonne National Laboratory, Lemont, 
 IL 60439, USA}
 \affiliation{Department of Physics, Northern Illinois University, DeKalb, IL 
 60115, USA}
\author{Omar Chmaissem}
 \affiliation{Materials Science Division, Argonne National Laboratory, Lemont, 
 IL 60439, USA}
 \affiliation{Department of Physics, Northern Illinois University, DeKalb, IL 
 60115, USA}
\author{Matthew Krogstad}
 \affiliation{Advanced Photon Source, Argonne National Laboratory, Lemont, IL 
 60439, USA}
\author{Daniel M. Pajerowski}
 \affiliation{Neutron Scattering Division, Oak Ridge National Laboratory, Oak 
 Ridge, Tennessee 37831, USA}
\author{Feng Ye}
 \affiliation{Neutron Scattering Division, Oak Ridge National Laboratory, Oak 
 Ridge, Tennessee 37831, USA}
\author{Duck Young Chung}
 \affiliation{Materials Science Division, Argonne National Laboratory, Lemont, 
 IL 60439, USA}
\author{Mercouri G. Kanatzidis}
 \affiliation{Materials Science Division, Argonne National Laboratory, Lemont, 
 IL 60439, USA}
 \affiliation{Department of Chemistry, Northwestern University, Evanston, IL, 
 60208, USA}
\author{Stephan Rosenkranz}
 \affiliation{Materials Science Division, Argonne National Laboratory, Lemont, 
 IL 60439, USA}
\author{Raymond Osborn}
 \affiliation{Materials Science Division, Argonne National Laboratory, Lemont, 
 IL 60439, USA}

\date{\today}

\begin{abstract}

Methods of elucidating the mechanisms of fast-ion conduction in solid-state
materials are pivotal for advancements in energy technologies such as batteries,
fuel cells, sensors, and supercapacitors. In this study, we examine the ionic
conduction pathways in single crystal SrCl$_2$, which is a fast-ion conductor
above 900~K, using four-dimensional Quasi-Elastic Neutron Scattering (4D-QENS).
We explore both coherent and incoherent neutron scattering at temperatures above
the transition temperature into the superionic phase to explore the correlated
motion of hopping anions. Refinements of the incoherent QENS yield residence
times and jump probabilities between lattice sites in good agreement with
previous studies, confirming that ionic hopping along nearest-neighbor
directions is the most probable conduction pathway. However, the coherent QENS
reveals evidence of de Gennes narrowing, indicating the importance of ionic
correlations in the conduction mechanism. This highlights the need for
improvements both in the theory of ionic transport in fluorite compounds and the
modeling of coherent 4D-QENS in single crystals.

\end{abstract}

\maketitle

\section{Introduction}
Ionic conductivity plays a critical role in a wide range of energy applications,
including batteries, fuel cells, sensors, and supercapacitors \cite{Wang2015}.
Fast-ion conductors, which exhibit high ionic mobility, are particularly
valuable for enabling efficient charge transport in these applications
\cite{Jun2024, Boyce1979}. Understanding the mechanisms of ionic conduction,
including the interplay between structural dynamics, correlated motion, and
defect-mediated pathways, is therefore pivotal for the design of next-generation
materials \cite{Funke1993, Boyce1979, Jun2024}. Exploring these mechanisms is
essential for advancing the design of safer, more efficient, and scalable energy
storage systems \cite{Wang2023, Hull2004, Boyce1979, Jun2024}.

In many solid state ionic conductors, mobile ions undergo random hopping
processes, moving between vacant sites created either by doping or thermally
activated Frenkel defects \cite{Hutchings1984,Hull2004}. However, there is
evidence in some materials of the importance of metastable interstitial sites
along the hopping pathway \cite{catlow1982,Tufail2023,Jun2024} or cooperative
ionic motion, where the movement of one ion facilitates the motion of nearby
ions by lowering the energy barrier for subsequent hops \cite{He2017, Jun2024}.
While simulations have highlighted the significance of ionic correlations in
conduction mechanisms, experimental techniques capable of directly detecting and
characterizing this phenomenon remain underdeveloped
\cite{Gillan1980b,Gillan1980a,Ishii2024,Dixon1980,Jun2024}.

One of the primary microscopic techniques for probing ionic dynamics in
solid-state conductors is Quasi-Elastic Neutron Scattering (QENS)
\cite{Funke1991, Springer2005, Schwaighofer2025}. QENS is most frequently used
to study the hopping processes of mobile ions in polycrystalline materials and
analyzed using jump diffusion models, such as the Chudley-Elliott model
\cite{Chudley1961}, to yield spherically-averaged values of hopping rates. There
are fewer single crystal studies, and these have been confined to measurements
along high-symmetry directions due to instrumental limitations
\cite{Hutchings1984, Hull2004}. However, the discovery of ionic conductors with
more complex hopping pathways \cite{Jun2024} and correlated ionic motion
\cite{Boivin2001,Abraham1990}, require these limitations to be overcome in order
to advance our understanding of ionic diffusion mechanisms in next-generation
conductors.

We propose that 4D-QENS offers a promising avenue for addressing this gap. Over
the past decade, it has become increasingly common to perform time-of-flight
inelastic neutron scattering with continuous or quasi-continuous sample rotation
to generate four-dimensional hyper-volumes of scattering in reciprocal space,
\textbf{Q}, and energy transfer, $\omega$ \cite{Ewings2016}. This allows the
energy dependence of scattering at constant-\textbf{Q} to be extracted, unlike
conventional time-of-flight spectroscopy measured at a fixed sample angle. So
far, these measurements have been used to generate 4D-S(\textbf{Q},$\omega$) for
the analysis of excitation spectra, providing methods of exploring, for example,
phonon anharmonicity or intra-band transitions in correlated electron systems
\cite{LaniganAtkins2021, Goremychkin2018}. However, in this report, we show that
it is just as valuable in the analysis of QENS measurements, allowing the
\textbf{Q}-dependence of quasi-elastic linewidths to be determined over
substantial reciprocal space volumes, not just high symmetry directions,
enabling more stringent comparisons with theoretical models. 

While conventional neutron scattering methods have been effective in analyzing
incoherent QENS in terms of jump diffusion models \cite{Hutchings1984},
measurements of coherent QENS have so far provided much more limited
information. Unlike incoherent scattering, which probes the hopping of
individual ions, coherent scattering can reveal the impact of ionic correlations
on the hopping process. The coherent QENS intensity, S$_{coh}$(\textbf{Q}),
measures two-particle correlations, which are sensitive to local relaxations
around both the hopping ions and the vacancies they may leave behind. On the
other hand, coherent QENS linewidths, $\Gamma_{coh}$(\textbf{Q}), are sensitive
to the lifetimes of these correlations. If the local relaxations are much faster
than the residence times, then it may be valid to analyze the coherent
linewidths in terms of jump diffusion models in the same way as incoherent
scattering \cite{Gillan1985}. However, the linewidths may also reveal a much
more complex interplay of time scales, especially when there are metastable
sites along the hopping pathway. It is well known from studies of liquids and
amorphous materials that there can be a narrowing of $\Gamma_{coh}$(\textbf{Q})
close to peaks in S$_{coh}$(\textbf{Q}), a feature known as de Gennes narrowing
\cite{DeGennes1959}, which reflects the increased stability of more probable
ionic correlations. There are far fewer observations of de Gennes narrowing in
crystalline materials, although it has been predicted to apply to a model of
dilute interstitials \cite{Sinha1988}. In this report, we propose that that the
observation of de Gennes narrowing in ionic conductors provides evidence of
metastable ionic configurations in the ionic hopping process.

\begin{figure}[htbp]
    \centering
    \includegraphics[width=\linewidth]{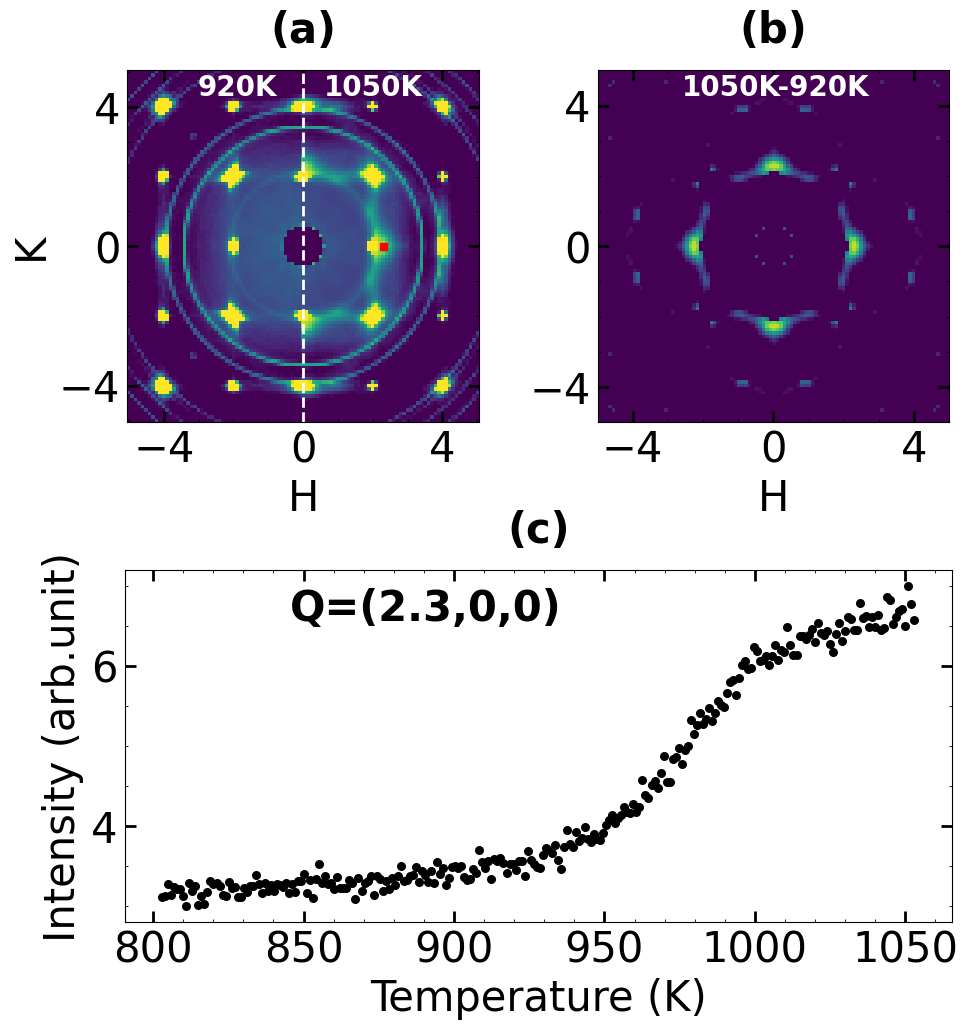}
    \caption{(a) Energy integrated neutron diffraction map in the [HK0]
    scattering plane of SrCl$_2$ measured on CORELLI at 920~K (left) and 1050~K
    (right). (b) the difference between the scattering intensity at 1050~K and
    920~K. (c) The temperature dependence of the diffuse scattering intensity at
    \textbf{Q}=(2.3,0,0), indicated by the red square..}
    \label{fig:elastic}
\end{figure}

This study uses 4D-QENS measurements to investigate ionic conductivity in the
superionic phase of a fluorite compound, SrCl$_2$, which has been investigated
using more conventional neutron scattering methods in the past
\cite{Hutchings1984, Hull2004}. Our results confirm the applicability of the
Chudley-Elliott jump diffusion model to describe the incoherent scattering of
hopping anions \cite{Chudley1961,Hutchings1984}, which dominates the scattering
at low \textbf{Q}, with improved accuracy because of the continuous
three-dimensional coverage in reciprocal space. However, this work provides
fundamentally new insights into the role of ionic correlations revealed by the
\textbf{Q}-dependent coherent linewidths, in particular the observation of de
Gennes narrowing, providing evidence of metastable ionic configurations during
the hopping process that were not predicted by previous theoretical studies
\cite{Dixon1980,Gillan1986a}. This demonstrates the power of 4D-QENS in
revealing details of ionic conduction mechanisms that are impossible to
determine by other experimental techniques.

\section{Experiment}
SrCl$_2$ was intensively studied by the Hutchings group over forty years ago,
along with a number of other fluorite compounds that exhibit fast-ion conduction
just below their melting temperature \cite{Hutchings1984}. SrCl$_2$ is ideal for
investigating ionic conductivity because strontium is a nearly perfect coherent
scatterer ($\sigma_{coh}=6.19$~barns, $\sigma_{incoh}=0.06$~barns), whereas
chlorine has strong coherent and incoherent cross sections
($\sigma_{coh}=11.53$~barns, $\sigma_{incoh}=5.3$~barns). In the fast-ion phase,
the strontium ions are considered to remain on the regular cation sites, so both
the coherent and incoherent scattering away from the Bragg peaks are dominated
by the mobile chlorine anions, whose diffusion generate the quasi-elastic energy
broadening.

The SrCl$_2$ crystals used in our experiments were grown using a commercial
reagent (BeanTown Chemical Corp., anhydrous beads, 99.999\% purity) without
additional purification. Ten grams of SrCl$_2$ were vacuum-sealed in a
carbon-coated silica tube with a conical tip. Crystal growth was carried out by
the electrodynamic gradient method using a 25-zone furnace. The temperatures of
the heating zones were controlled by a program with 12 segments. The temperature
interval near the crystallization point was 15\textdegree C/cm and the
temperature translation rate is 3\textdegree C/h. The resulting crystal was
clean and transparent in most part, while the upper fraction was slightly
cloudy. For this study, only the high-quality crystals from the bottom section
were used.

Measurements of the diffuse scattering in SrCl$_2$ were performed on the CORELLI
instrument at the Spallation Neutron Source \cite{Ye2018}. This allows the
energy-integrated diffuse scattering, S(\textbf{Q}) to be determined over a
large reciprocal space volume as a function of temperature. The sample, which
was 15x5x5~mm$^3$ in size, was mounted in a high-temperature furnace for
measurements between 800~K and 1100~K. In agreement with earlier studies
\cite{Hutchings1984}, the most significant diffuse scattering is seen beyond the
(200) Bragg peak, with an intensity that grows strongly on entering the
superionic phase (Fig. 1).

The QENS measurements were performed on the Cold Neutron Chopper Spectrometer
(CNCS) at the Spallation Neutron Source \cite{Ehlers2016}. The sample was
mounted in a furnace on a rotating stage that allowed scans over 360$^\circ$
range in steps of 1$^\circ$. This rotation allowed the reconstruction of a full
4D volume of data covering \textbf{Q}-space and energy transfer. Both the
CORELLI and CNCS data were reduced and symmetrized using the Mantid software
suite \cite{Savici2022,Arnold2014}.

For reasons to be discussed in the next section, we performed the CNCS
measurements with two different incident energies. By using an incident energy
of 1.55~meV on CNCS, we were able to measure QENS up to $|$Q$|$ = 1.85
\AA$^{-1}$, with an energy resolution of 15.4~$\mu$eV, estimated from scans
measured well below the super-ionic transition temperature at 850K (Fig. 2a).
With an incident energy of 3.6~meV, we extended the Q-coverage up to $|$Q$|$ =
2.5 \AA$^{-1}$, with an energy resolution of 49.8~$\mu$eV (Fig. 2b). 

\begin{figure}[htbp]
    \centering
    \includegraphics[width=\linewidth]{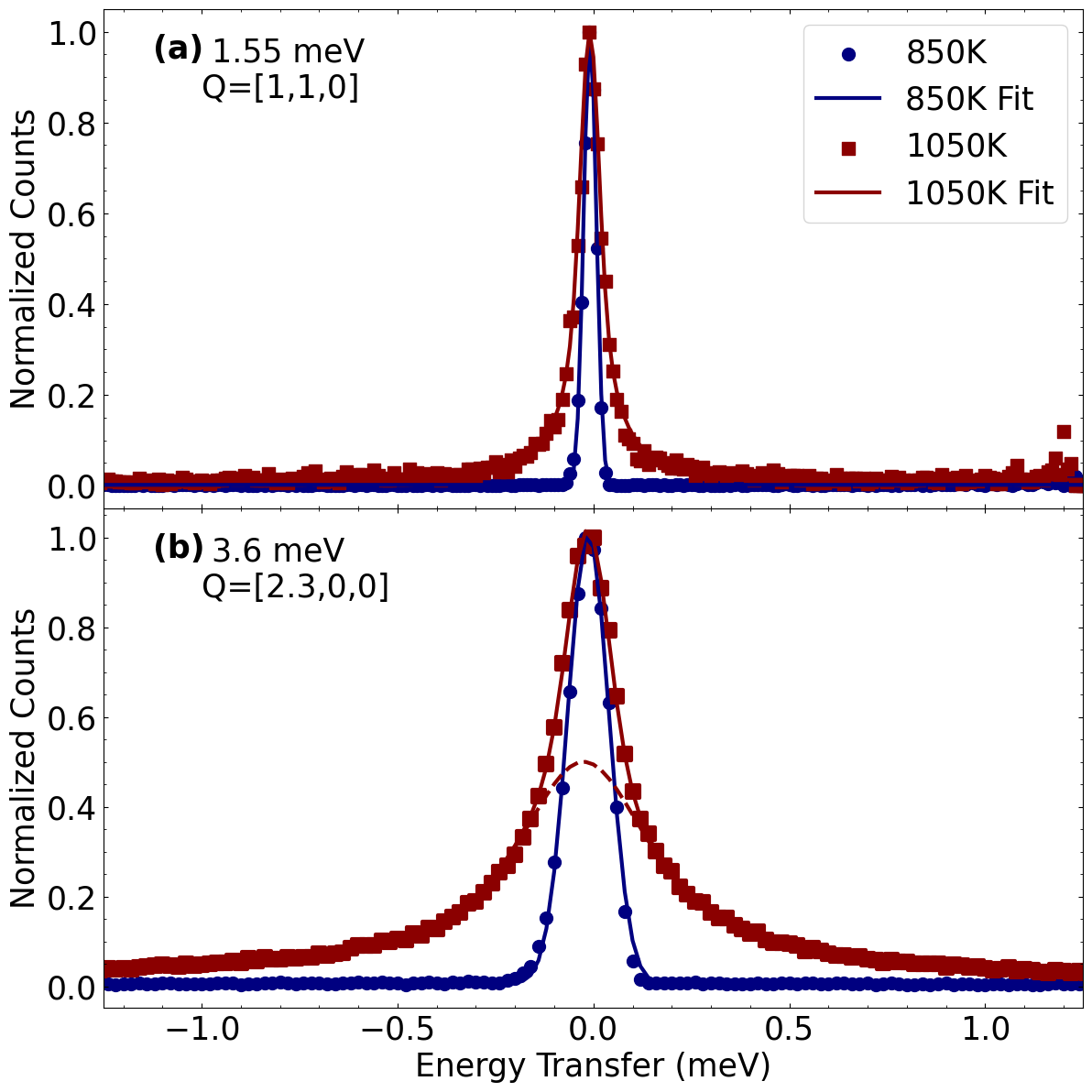}
    \caption{(a) Energy distributions for incident energy 1.55 meV measured at
    the \textbf{Q}=[1,1,0] position at 1050K, above the ionic conduction
    transition, and at 850K, below the ionic transition. This can be compared
    with (b) 3.6 meV incident energy measurements of \textbf{Q}=[2.3,0,0], the
    region highlighted by a red box in in Figure \ref{fig:elastic} (a), showing
    larger energy broadening due to the coherent scattering represented by a
    dashed line.}
    \label{fig:distribution}
\end{figure}

\section{Results}
The measurements performed on CORELLI confirm that there is a growth in coherent
diffuse scattering associated with the superionic phase above $\sim$920~K (Fig.
1). The 4D-QENS measurements on CNCS were therefore taken at 1050~K, about 100~K
below the melting temperature. The 4D grid of scattering intensity following
data reduction using the Mantid package was divided into cubes of 0.1 reciprocal
lattice units (r.l.u.) in order to model the quasi-elastic broadening at each
\textbf{Q}. S(\textbf{Q},$\omega$) consists of contributions from elastic,
quasi-elastic, and inelastic components, as well as instrumental background.
\cite{Boothroyd2020, Springer2005, Funke1991}:
\begin{equation}
\begin{split}
    \mathbf{S}(\textbf{Q},\omega) = & [A_0(\mathbf{Q})\delta(\omega) + A_{1}
    (\mathbf{Q}) S_{incoh}(\mathbf{Q},\omega) \\&+ A_{2}(\textbf{Q}) S_{coh}
    (\mathbf{Q},\omega)+B(\textbf{Q},\omega)] \circledast R(\mathbf{Q},\omega)
\end{split}
\end{equation}
In this equation, $A_0$(\textbf{Q}) is the intensity of purely elastic
scattering mostly from instrumental backgrounds as well as very weak strontium
incoherent scattering. $A_1$(\textbf{Q}) and $A_2$(\textbf{Q}) are the
incoherent and coherent scattering contributions, respectively, from the
chlorine ions. Both are broadened in energy by the chlorine ion diffusion,
modeled by Lorentzian lineshapes of differing half-widths,
$\Gamma_{incoh}(\mathbf{Q})$ and $\Gamma_{coh}(\mathbf{Q})$, respectively. The
final contribution, $B$(\textbf{Q},$\omega$), represents inelastic
contributions, mostly from phonon and multiphonon scattering, but in practice,
it was too small at such low energies and wavevectors to be included in our
analysis apart from a constant background. The elastic and quasi-elastic
components are all broadened by a Gaussian resolution function,
$R$(\textbf{Q},$\omega$), whose energy widths at each incident energy were fixed
by measurements at 850~K.

\subsection{Incoherent Scattering}
When analyzing 4D-QENS, it is not possible \textit{a priori} to separate the
coherent and incoherent components in the scattering cross section. However, as
discussed by Hutchings \textit{et al} \cite{Hutchings1984}, the energy
broadening of incoherent scattering is predicted to be much smaller than
coherent scattering. This is because incoherent energy broadening results from
the time-dependence of the self correlation function, so it reflects the average
residence times of all the anions. On the other hand, the energy broadening of
coherent diffuse scattering results from the time-dependence of two-particle
correlation functions involving defects in the anion sublattice, either
vacancies or interstitials or both, along with relaxations of neighboring ions,
during the hopping process. If, for example, only 10\% of the anion sites are
involved in hopping at any one time, \textit{i.e.}, they host a vacancy or
interstitial anion, the energy scale of coherent QENS will be an
order-of-magnitude larger than incoherent QENS (although both are
\textbf{Q}-dependent, so the ratio is not precise).

\begin{figure}[htbp]
    \centering
    \includegraphics[width=\linewidth]{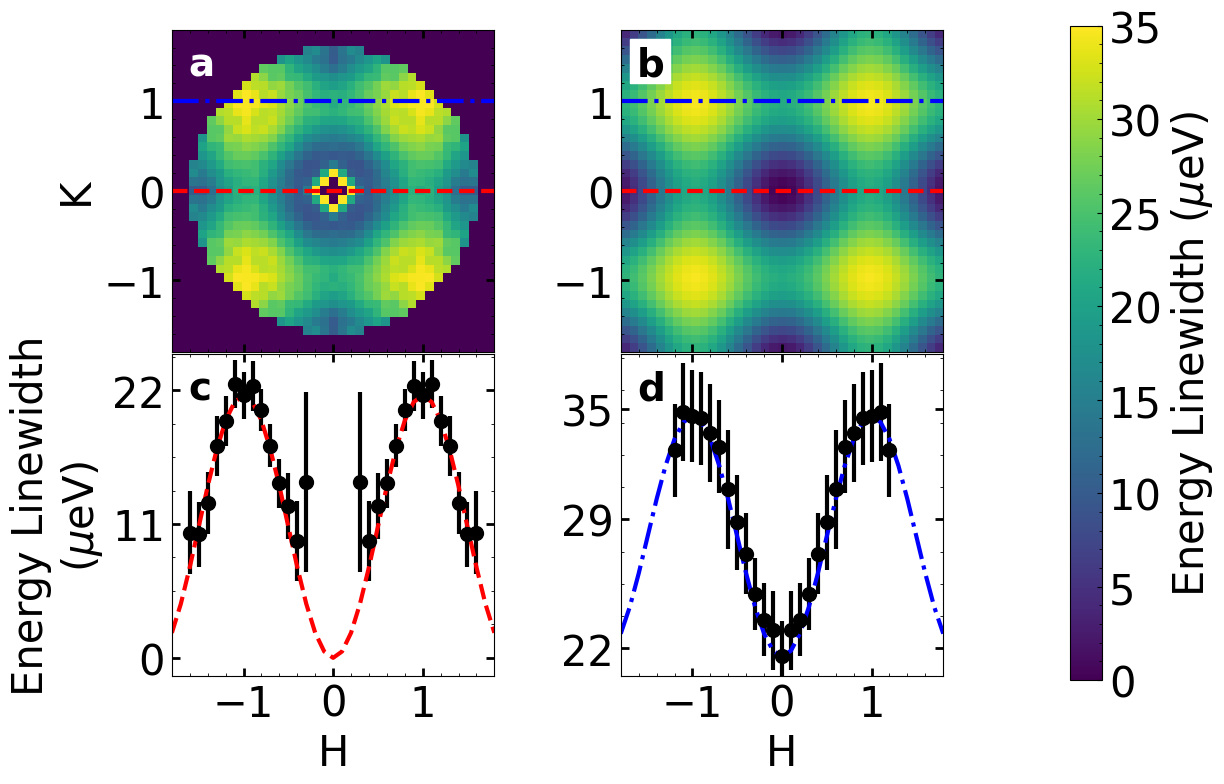}
    \caption{ Plots of the L = 0 plane of (a) the incoherent linewidths measured
    using incident energy E$_i$~=~1.55~meV compared to (b) the Chudley-Elliott
    model with parameters p$_1$~=~0.754~$\pm$~0.002, p$_2$~=~0.746~$\pm$~0.002,
    and $\tau_{inc}$~=~25.36~x~10$^{-12}$ demonstrating general visual fit. One
    dimensional line-cuts taken at (c)~K~=~0 and (d)~K~=~1 further illustrate
    the quality of fit between the measured linewidths (points) and the model
    (dashed line).}
    \label{fig:155linewidths}
\end{figure}

As shown in the next section, the coherent diffuse scattering is weak at low
\textbf{Q}, below the (200) Bragg peak, so the high resolution measurements were
dominated by incoherent scattering. An example of scattering above and below the
transition temperature measured with an incident energy of 1.55 meV can be seen
in Fig.~\ref{fig:distribution}, showing a clear broadening at 1050~K, which is
absent at 850~K. Using the least-squares method \cite{Newville2024}, we fit such
data to the sum of a Voigt and a Gaussian function, with the Gaussian standard
deviation of both functions constrained to be equal to the estimated resolution.
By automating such fits over the measured volume in reciprocal space, we were
able to generate a 3D map of $\Gamma_{incoh}(\mathbf{Q})$, without being
confined to high symmetry directions. The resulting Q-dependence of the
incoherent linewidth in the [h,k,0] plane is shown in
Fig.~\ref{fig:155linewidths}(a). 

These Q-maps were then fit in three dimensions to the Chudley-Elliott (C.E.)
jump diffusion model \cite{Chudley1961}, which assumes that the ions jump
instantaneously between lattice sites. The model predicts that the linewidths
will increase quadratically at low-Q, but then fall to zero at all the
wavevectors corresponding to the hopping sublattice. \cite{Dickens1983,
Rowe1971}: 
\begin{equation}\label{CEModel}
    \Gamma_{incoh} (Q) = \frac{1}{2 \pi \tau_{incoh}} \sum _i{ \frac{p_i}{n_i} 
    \left(\sum _j{[1-exp(-i\mathbf{Q} \cdot \mathbf{r}_{i j})]}\right)} 
\end{equation}
In this equation, $\tau_{incoh}$ is the mean residence time of anions between
hops, $p_i$ is the probability of hopping to the $i$th nearest-neighbor, $n_i$
is the number of equivalent $i$th nearest neighbors, and $r_{i j}$ is the
interatomic vector connecting sites $i$ and $j$. 

As shown in Figure  \ref{fig:155linewidths}, our data strongly supports the C.E.
model predictions, with four lobes centered symmetrically around Q=0. The
results of this analysis are compared to the results in (a) using 1-D line cuts
along the [H,0,0] and [H,1,0] directions, as presented in (c) and (d)
respectively. Based on our fits , we find that $p_1=0.754\pm~0.002$,
$p_2=0.246\pm0.002$, and $\tau_{incoh}=25.36\times10^{-12}$~s. These values are
consistent with previous experimental estimates in SrCl$_2$
\cite{Dickens1983,Hutchings1984,Rowe1971}.

We attempted to include additional hopping vectors, but the values of $p_n$ were
0 within errors for any $n\ge~3$. We also checked if the anions might hop to and
from `interstitial' sites, such as those predicted by the Hutchings group
\cite{Hutchings1984}, with similar results. It is worth noting that such
additional hops can substantially distort the shape of the four lobes shown in
Figure \ref{fig:155linewidths}a, so we believe the conclusion that only nearest
and next-nearest neighbor hops have a significant probability is more robust in
the 4D-QENS analysis than in previous measurements that were confined to a few
high-symmetry directions \cite{Dickens1983}.

\subsection{Coherent Scattering}
\begin{figure*}[ht]
    \centering
    \includegraphics[width=\linewidth]{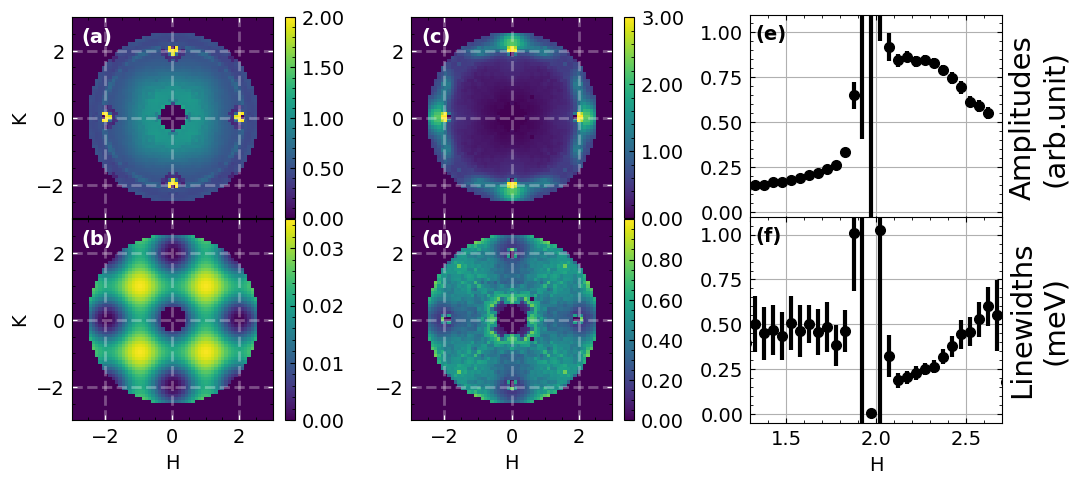}
    \caption{ Component results of fits of the measurements taken at
    E$_i$~=~3.6~meV are presented. Shown are the L~=~0 planes of (a) incoherent
    structure factor, (b) incoherent linewidths generated using the
    Chudley-Elliott model found at lower energies, (c) coherent structure
    factor, (d) coherent linewidths. Additional fits were performed with binning
    along the H direction reduced, with the coherent amplitudes of these fits
    shown in (e) and the linewidths shown in (f). }
    \label{fig:36symm}
\end{figure*}

We used the results of the C.E. model refinement to constrain our analysis of
the 3.6~meV data at 1050~K, in which there are significant contributions from
coherent scattering. With the extended energy range, we fit the data to the sum
of a Gaussian function, from purely elastic scattering, and two Voigt functions,
representing the incoherent and coherent quasi-elastic scattering, respectively.
However, the parameters of the incoherent Voigt function linewidths were fixed
to those predicted by the C.E. model, allowing the coherent contributions to be
determined without excessive parameter correlations. The fit results are shown
in Fig.~\ref{fig:36symm}, with (a) showing the incoherent structure factor, (b)
the incoherent linewidths, (c) the coherent structure factor, and (d) the
coherent linewidths. Note that the incoherent linewidths shown in
Fig.~\ref{fig:36symm}b are calculations based on the C.E. model derived from the
1.55~meV data, whereas Fig.~\ref{fig:36symm}a are fits, whose Q-dependence is
consistent with a monotonic fall in the integrated incoherent intensity from the
Debye-Waller factor.

Fig.~\ref{fig:36symm}c shows the integrated coherent intensity, which is
consistent with the CORELLI data shown in Fig.~\ref{fig:elastic}b, as well as
previous neutron measurements \cite{Hutchings1984}. However, the most
interesting result is shown in Fig.~\ref{fig:36symm}d. Although the coherent
intensity is weak at wavevectors below the (200) Bragg peaks, it is nevertheless
reliably determined from the fits, allowing the coherent linewidths to be
examined over the entire volume in reciprocal space for the first time. It is
clear that the $\Gamma_{coh}$ bears no resemblance to $\Gamma_{incoh}$, which
falls to zero symmetrically around \textbf{Q} = (200). Instead, there is a
strong asymmetry, with a significant dip in the linewidths at wavevectors beyond
the (200) Bragg peaks. Line cuts along the longitudinal direction in Q show a
broad maximum in coherent diffuse intensity at $\mathrm{H}\gtrsim 2.2$
(Fig.~\ref{fig:36symm}e), correlated with a comparable reduction in the
linewidths (Fig.~\ref{fig:36symm}f), although the minimum wavevector is not as
clearly defined in our analysis. This is characteristic of de Gennes narrowing
\cite{DeGennes1959}, in which the linewidths are minimized at wavevectors
corresponding to maxima in the diffuse S$_{coh}$(\textbf{Q}). This will be
discussed in the next section.

\section{Discussion}
Examples of de Gennes narrowing have mostly been confined to liquids and
amorphous materials, including molten SrCl$_2$ \cite{Margaca1984}, but the only
microscopic derivation in crystalline materials has been provided by Sinha and
Ross \cite{Sinha1988}, who modeled the solid state diffusion of a dilute
concentration of hydrogen interstitials. Using linear response theory, they
predicted that the coherent linewidths could be derived from the incoherent
linewidths by the following equation \cite{Sinha1988, Springer2005}:
\begin{equation}\label{eq:sinha}
    \Gamma_{coh}(\mathbf{Q}) = \frac{c(1-c)\Gamma_{incoh}(\mathbf{Q})}{S_{coh}
    (\mathbf{Q})}
\end{equation}
where \textit{c} is the concentration of interstitials. However, equation
\ref{eq:sinha} is inconsistent with the global \textbf{Q}-dependence shown in
Fig.~\ref{fig:36symm}b, c, and d.  We suggest that this is due to the neglect of
ionic correlations in the dilute limit \cite{Schwaighofer2025}. Nevertheless the
observed coincidence of a maximum in $S_{coh}(\mathbf{Q})$ with a minimum in
$\Gamma_{coh}(\mathbf{Q})$ is consistent with the intuitive interpretation of de
Gennes narrowing that more probable interionic correlations are associated with
an increased lifetime of the corresponding ionic configurations.

The reason for the increased lifetimes goes to the heart of the ionic conduction
mechanism. In the Hutchings study \cite{Hutchings1984}, the diffuse scattering
was modeled in terms of anion clusters, comprising an interstitial anion,
located half-way between the regular anion sites but shifted towards a
neighboring empty cube center, as well as a number of neighboring anions
relaxing away from it. Whether these clusters arose from thermally activated
Frenkel defects that were not associated with vacancy hopping or were part of
the hopping process itself could not be determined from their measurements. This
cluster model was, however, challenged by molecular dynamics (MD) simulations
based on realistic pair potentials \cite{Gillan1986a, Gillan1986b}, which were
able to reproduce the measured S(\textbf{Q}) from an average over the
instantaneous ionic configurations, without seeing any evidence of the proposed
interstitial clusters in the anion trajectories. Instead, Gillan proposed that
anions randomly hopped directly from one regular site to another
\cite{Gillan1980b}, producing an equal number of vacancies and doubly-occupied
sites (which Gillan called ``interstitial" sites), where the occupation number
is defined by a large sphere centered on each site \cite{Gillan1986b}. According
to the MD simulations, the strong peaks in diffuse scattering beyond the (200)
Bragg peaks are produced predominantly by correlations between anions on the
doubly-occupied sites, both of which are presumably strongly relaxed from the
site center. On the other hand, relaxations of neighboring ions around the
vacancy sites mostly contribute to S$_{coh}$(\textbf{Q}) in regions where the
scattering is relatively weak.

The strong \textbf{Q} dependence of the coherent linewidths supports the
conclusion that the peak in S$_{coh}$(\textbf{Q}) is associated with the hopping
process itself, since thermal activation of isolated clusters would produce a
\textbf{Q}-independent linewidth. Furthermore, the observation of de Gennes
narrowing implies that there is a finite probability for anions to be trapped on
doubly-occupied sites until one of them is able to hop to a neighboring vacancy
site. The MD simulations show evidence of chains of correlated hops consistent
with this scenario. With a dilute concentration of thermally activated
vacancies, the residence time of doubly-occupied sites could be substantially
longer than the residence time of vacancies determined from the coherent
broadening below the (200) Bragg peaks (Fig.~\ref{fig:36symm}f).

\section{Conclusion}
Our study shows that 4D-QENS, a variation of QENS, in which single crystal
samples are continuously rotated through 360\textdegree\ while collecting data
as a function of neutron time-of-flight, is a viable method for characterizing
ionic hopping mechanisms in fast-ion conductors such as SrCl$_2$. It allows the
reconstruction of S(\textbf{Q},$\omega$) in four dimensions to facilitate
comparisons with models of both incoherent and coherent cross sections as a
function of wavevector transfer. The high-resolution measurements of incoherent
linewidths presented here largely confirm the results of earlier investigations
of the incoherent scattering, which were analyzed in terms of the
Chudley-Elliott jump diffusion model \cite{Hutchings1984}. However, the previous
measurements were confined to high-symmetry directions in reciprocal space,
which would not have been as sensitive to additional hopping terms, such as to
hypothetical interstitial sites. Conclusions that the only significant hopping
probabilities are between nearest-neighbor and next-nearest-neighbor anion sites
are therefore more robust in the present study where off-symmetry directions are
included.

Our measurements of the coherent diffuse scattering are also broadly consistent
with the earlier studies, with the most significant intensity observed at
wavevectors above the (200) Bragg peaks. However, 4D-QENS has elucidated the
\textbf{Q}-dependence of the linewidths in unprecedented detail, covering a
broad dynamic range of diffuse scattering intensities in a fine \textbf{Q}-mesh.
This has provided evidence of de Gennes narrowing, resulting from the enhanced
lifetimes of defects, in which two anions are centered on the same site. We hope
that this work will stimulate an extension of the theory of de Gennes narrowing
in crystalline materials to systems, in which there are strong interactions
between the diffusing ions, as well as future experiments utilizing the 4D-QENS
method to investigate ionic diffusion pathways in more complex ionic conductors.

\begin{acknowledgments}
This work was supported by the US Department of Energy, Office of Science,
Office of Basic Energy Sciences, Materials Science and Engineering Division and
Scientific User Facilities Division. This research used resources at the
Spallation Neutron Source, a DOE Office of Science User Facility operated by the
Oak Ridge National Laboratory. Beam time was allocated at CNCS under proposal
IPTS-26314, and additional beam time was allocated at CORELLI under proposal
IPTS-26395.
\end{acknowledgments}

\bibliography{references}

\end{document}